\input phyzzx.tex
\hoffset=10mm
\vsize=26cm
\date={NUP-A-98-6}
\titlepage
\title {\bf Space-Time Quantization and Matrix Model}
\author {Sho TANAKA\footnote{*}{ Em. Professor of Kyoto University and 
Associate Member of Atomic Energy Research Institute, Nihon University.}}
\address {Kurodani 33-4, Sakyo-ku, Kyoto 606-8331, Japan}
\abstract{In order to get the general framework describing a nonlocalizable 
object beyond the bilocal field theory early proposed by Markov
 and Yukawa, 
the quantization of space-time is reconsidered and further developed. 
Space-time quantities are there not only noncommutative with $U$-field 
describing the nonlocalizable 
object, as in the bilocal field theory, but also become noncommutative 
among themselves. Under the $U$-field representation, where the basis
vectors of representation are chosen to be eigenvectors of operator $U$,
space-time quantities get a matrix representation of infinite
dimension in general. Field equation is considered, which determines
the relation
between space-time quantities and $U$-field. The possible inner relation 
between the recent topics of matrix model in superstring  theory and 
the present approach is discussed.}

PACS 12.90 - Miscellaneous theoretical ideas and models.
\endpage
\chapter{Introduction}
\def\yukawa{H.~Yukawa, Prog.~Theor.~Phys.~{\bf 2} (1947), 209.}
\def\markov{M.~A.~Markov, Jour.~Phys.~{\bf 2} (1940), 453.}
 Fifty years ago, Yukawa\ref\yukawa proposed the nonlocal field theory, 
according to the preceding idea by Markov,\ref\markov 
discarding the restriction that the field quantities are simply the
functions of space-time coordinates. The field concept is there
extended so that the field, let us call it $U$ in general, is no longer
commutative with the space-time coordinates;
$$ 
    [{\rm A}] \qquad \qquad \qquad       [ U, x_\mu ] \neq 0.
\eqn\eqA
$$     
The motivation for this attempt was, on the one hand, to remove the
long-pending problem  of the so-called divergence difficulty inherent
in the local field 
theory, and on the other hand, to give the unified description of
elementary particles, in the face with the unexpected discovery of the
mu-meson
in the development of the meson theory, i.e., the first appearance of
generation structure of elementary particles.  
It was expected that the condition \eqA\ will enable us to provide the
space-time uncertainty or the nonlocal extension to the field $U$,
beyond the 
local field concept based on the point particle model of elementary
particles.

In fact, Yukawa showed that the $U$-field obeying \eqA\ becomes the
so-called bilocal field described by the two-point function as
$$
      \langle x'_\mu \vert U \vert x''_\mu \rangle \equiv U(x'_\mu,x''_\mu)
      \equiv   U(X_\mu,r_\mu),
\eqn\eqB    
$$   
with $  X_\mu \equiv (x''_\mu + x'_\mu) / 2 $ and $ r_\mu \equiv x''_\mu - 
x'_\mu$,
by taking the space-time coordinate representation where the basis vectors 
of representation consist of states with the definite values of four
space-time coordinates like $ \vert x'_\mu \rangle $ or $ \vert
x''_\mu \rangle.$ He 
further asserted that the different modes of functional behavior with
respect to the new internal degrees of freedom of the relative
coordinate $ r_\mu $  have the possibility of explaining the origin of
the variety of elementary particles with  different masses  and
(integer) spins, although he did not succeed 
to find  the satisfactory way of introducing the interactions among
bilocal fields.

In the present paper, we start with presenting an elementary, but very serious
question why the proposition [A] in \eqA\  which asserts in the general
form the non-commutativity of the field $U$ with the space-time
coordinates remains to provide 
only the bilocal field suitable for the description of the system like
the two-quark system, but not the multilocal fields. In fact, the
latter fields which 
correspond to the existing multi-quark systems had to be introduced
{\it by hand} 
merely as the formal extension of the bilocal field. We wonder why 
the proposition \eqA\ does not cover the extended objects more in
general such as string, membrane and so forth.   

The answer, however, seems  to be rather simple. In fact, it turns
out that 
from the assumption \eqA\ one arrives at the bilocal field immediately
after one takes the space-time coordinate representation, where the
four space-time 
coordinates are presumed to be commutative with each other and  all
described by
diagonal matrices at the same time. Therefore, in order to go over the
limit of
the bilocal field under the general proposition \eqA , let us  consider the 
possibility that 
space-time coordinates are noncommutative:
$$
     [{\rm B}]\qquad \qquad \qquad      [ x_\mu, x_\nu ] \neq o,
\eqn\eqC
$$
which clearly makes impossible to take the space-time coordinate
representation naively as was done in \eqB.

In the present paper, we wish to investigate the general framework to
describe the extended object  under the propositions [A] in \eqA\ and
[B] in \eqC. The 
latter proposition [B] is  well-known in the old attempts of the
space-time quantization.  Snyder\ref\snyder once proposed the idea of
``Quantized Space-Time", 
\def\snyder{H.~S.~Snyder, Phys.~Rev.~{\bf 71} (1947), 38.}
nearly at the same 
time when  the nonlocal field theory  was proposed by Yukawa. Later on, 
Yang\ref\yang gave the special attention to the Snyder's work and
presented an argument 
\def\yang{C.~N.~Yang, Proc.~of International Conf.~on Elementary
  Particles, 1965 Kyoto, pp 322-323.}
to modify it. 

In the next section  2, we reconsider the Snyder's work as one
possible realization of the proposition [B], by supplementing the
argument given by Yang. In the section 3, we study the propositions
[A] and [B] together and present, instead of the space-time coordinate
representation, newly  $U$-field representation as a realization of
algebra obeying [A] and [B].  In this representation, $U$-field becomes
a diagonal matrix, while  space-time quantities tend to the infinite
dimensional matrices in general. We further investigate the field
equation, which 
relates $U$-field to the quantized space-time. The final section 4 is
devoted to discussions in which we consider the possible relation
between the present approach and the matrix model proposed 
in the recent string theory.   
         
\chapter{Space-Time Quantization}

As was mentioned in Introduction, with the aim of getting the finite
and Lorentz-invariant theory, Snyder \refmark{3} presumed the
five-dimensional (de Sitter) space in the background of real
space-time and attempted to express space-time quantities as linear
differential 
operators on the former space, which are noncommutative with each
other. Nearly two decades later, Yang \refmark{4} gave special
attention to the Snyder's work in the point that space coordinates
have discrete eigenvalues and time coordinate continuous, 
but the theory is Lorentz-invariant.

In what follows, let us reconsider the Snyder's theory in
supplementing Yang's
argument. Yang proposed to modify the background space, from the original
five-dimensional (de Sitter) space to six-dimensional one 
$(\xi_0, \xi_1,\xi_2,\xi_3,\eta,\tau)$, constrained as 
$$
        - \xi_0^2  + \xi_1^2 + \xi_2^2 + \xi_3^2 +\eta^2 +\tau^2 =
        {\rm const.}.
\eqn\eqD
$$
Space-time coordinates are defined;
$$
    X_i = i(\xi_i\partder{}{\eta}-\eta\partder{}{\xi_i}),
 \eqn\eqE
$$
$$
    X_0 = i(\xi_0\partder{}{\eta}+\eta\partder{}{\xi_0})
\eqn\eqF
$$
in the same way as Snyder. The four-momentum coordinates are similarly
defined, replacing $\eta$ with $\tau$ in the above equations as
$$
    P_i = -i(\xi_i\partder{}{\tau} -\tau \partder{}{\xi_i}),
    \eqn\eqG
$$
$$
    P_0 = -i(\xi_0\partder{}{\tau}+\tau\partder{}{\xi_0}),
    \eqn\eqH
$$
differently from Snyder. Actually, Snyder considered them, not as
quantized ones, but in the following way; $ P_i=\xi_i/\eta $ and $
P_0=\xi_0/\eta$.             
In the above expressions, one easily
finds that both $X_i$ and $ P_i (i=1,2,3))$ have 
integer eigenvalues because they are angular momentum operators with respect 
to $ \xi_i-\eta$ and $\xi_i-\tau$ planes, respectively.   

At this point, one wonders how the Lorentz invariance is guaranteed in
conformity with the discrete eigenvalues of each i-th component as
well as the continuous
eigenvalues of 0-th component.  The clue to this question lies in the fact
that they are all noncommutative quantities. This thing is suggested from the 
familiar quantum-mechanical angular momenta, which have discrete eigenvalues,
but remain to be of rotation-invariant character.

In fact, one can confirm the above fact by calculating the following
commutation relations;
$$
      [ X_i,X_j]=-iL_{ij},
      \eqn\eqI
$$
$$    [ X_i,X_0]=iM_i,
\eqn\eqJ
$$
and
$$
      [P_i,P_j]=-iL_{ij},
      \eqn\eqK   
$$
$$    
      [P_i,P_0]=iM_i,
      \eqn\eqL
$$
with $L_{ij}$ and $M_i$ defined by
$$
      L_{ij} = i(\xi_i \partder{}{\xi_j}-\xi_j\partder{}{\xi_i}),
      \eqn\eqM
$$
$$
      M_i=i(\xi_0\partder{}{\xi_i}+\xi_i \partder{}{\xi_0}).
      \eqn\eqN
$$
Further,
$$
      [X_i,P_j]=i\delta_{ij}N,
      \eqn\eqO
$$
$$
      [ X_0,P_0]=-iN,
      \eqn\eqP
$$
$$
      [X_i,P_0]=[P_i,X_0]=0
      \eqn\Q
$$      
with N defined by
$$
      N=i(\eta \partder{}{\tau}-\tau \partder{}{\eta}).
      \eqn\eqR
$$            

It is quite interesting here to note that the operator $N$, i.e.,
angular momentum with respect to $\eta-\tau$ plane, concerns the
origin of Planck constant $h$ in quantum mechanics, under the suitable
choice of scale units of $X_{i,0}$ and $P_{i,0}$ such as Planck
length, which were omitted so far. Furthermore, one notices that $N$
plays the role of junction between $X$ and $P$;
$$
    [X_i,N]=-iP_i,\qquad \qquad
      \cdots
    \eqn\eqS
$$

With respect to $ L_{ij}$ and $M_i$, one finds that they are nothing
but the six generators of Lorentz transformation, which constitute the
well-known Lorentz- Algebra with space-time quantities. It turns out
that the transformation corresponds to the special transformation in
the six-dimensional space with $\eta$ and $\tau$ fixed in \eqD and
guarantees the Lorentz invariance of the present theory. 

In closing this section, one should remark that the fifteen operators 
$$
{\cal R}_{15}\equiv(X_\mu,P_\mu;L_{ij},M_i;N)
\eqn\eqT
$$
with $\mu=(i,0)$, constitute as a whole a Lie ring, which characterizes  
the structure of our quantized space-time.   

\chapter{$U$-field Representation and Matrix Model of Space-Time}

In the preceding section, we have studied the modified Snyder theory as one 
interesting model of space-time quantization according to the
proposition [B] stated
in Introduction. In the present section, we consider the proposition [A],
i.e., the noncommutative structure of field $U$ with space-time quantities. 
If we assume, for the sake of explanation, the space-time structure to be 
expressed by ${ \cal R}_{15}$, \eqA\ becomes  written more
explicitly as\foot{We do not exclude the possibility that some ones of 
${\cal R}_{15}$, for instance, $ N $, are commutative with $U$.}
$$
        [{\rm A}'] \qquad \qquad \qquad   [ U, {\cal R}_{15}] \neq 0.
           \eqn\eqU 
$$          

One sees, however, that the above equation by itself is insufficient
to qualify the operator $U$, which is expected as a unified field
describing 
ultimately all the elementary particles and fundamental forces, as the
present 
string field theory aims. In the conventional field theory, the field
equation plays the role of determining the relation between the field
and space-time structure, although the space-time structure is already
fixed as given , except for the gravitational equation. In the present
case, it is desirable that the quantized space-time structure is
determined simultaneously with $U$, as in general theory of relativity.

Before entering into the discussion of this problem, it is 
important to notice the general feature characteristic of the
proposition [A'].
As was remarked in Introduction, it is now impossible to take the space-time 
coordinate representation to express the noncommutative relations such
as ${\cal R}_{15} $. Therefore, instead of this, we take a novel
representation, let us call it $U$-field representation, in which the
basis vectors of representation are chosen to be eigenstates of the
operator $U$. Let us assume that 
eigenvalues are discrete, for the sake of simplicity, i.e.,
$$
            U \vert n \rangle = u_n \vert n \rangle.
                    \eqn\eqV
$$
$U$ is represented now by a diagonal matrix of infinite-dimension, in general;
$$
            U = {\rm diag} (u_1,u_2,u_3, \cdots).
            \eqn\eqW
$$

Here the (ortho-normal) eigenvector $\vert n \rangle  (n=1,2,3,
\cdots) $ with eigenvalue 
$u_n$  may be imagined to describe the n-th excitation mode of $U$, whose 
space-time structure becomes explicit through the matrix representation of 
the space-time Algebra such as ${\cal R}_{15}$. Each space component $X_i$,
for instance, is represented by the matrix          
$$
        X_i = \big\lbrack \quad X_i^{n'n''} \quad \big\rbrack
        \eqn\eqX
$$
with the matrix element
$$      X_i^{n'n''} \equiv \langle n' \vert X_i \vert n''\rangle.
\eqn\eqY
$$

The diagonal matrix element $\langle n \vert X_i \vert n \rangle$ must denote
the space coordinate of the n-th excitation mode of $U$ with fluctuation 
$\Delta X_i^{(n)}$ defined through
$$
        (\Delta X_i^{(n)})^2 \equiv \langle n \vert( X_i - X_i^{(nn)})^2
        \vert n \rangle=\sum_{n'\neq n} \vert X_i^{nn'} \vert^2.
        \eqn\eqZ
$$        

Now we are in a position to consider the field equation of $U$, which
more explicitly relates $U$ to the space-time structure, as stated
above. Unfortunately, we have no reliable principle to determine the
form  of the 
equation, but the past nonlocal field theories such as the bilocal or
the string field theory, though they are not necessarily based on the
quantized 
space-time, seem strongly to suggest the following form, by neglecting
the interaction terms;
$$
      [ P_\mu,[ P_\mu,U ]] + {\cal M}^2 U =0.
      \eqn\eqAA
$$      
With respect to the mass-squared operator ${\cal M}^2$, string theory
suggests  that it 
involves the angular momentum in the Lorentz-invariant form
$$ 
          {L_{ij}}^2-{M_i}^2 (= {[X_\mu,X_\nu]}^2).
          \eqn\eqAB
$$
Furthermore, if we want to get the half-integer spin mode of $U$-field,
it becomes necessary to introduce relativistic spin-angular momentum
quantities beside $L_{i,j}, M_i$, which are known to be described in
terms two kinds of Euler's angles(left and right)\ref\bohm and serve
to linearize the field equation as Dirac equation.
\def\bohm{D.~Bohm, P.~Hillion, T.~Takabayasi and J.~P.~Vigier,
  Prog.~Theor.~Phys. {\bf23} (1960), 496; \nextline
  S.~Tanaka, Prog.~Theor.~Phys.~Suppl.~No.~{\bf67} (1979), 282.}
Needless to say, the above field equation can be transformed into the
matrix form, which works to constrain the matrix elements of
space-time operators as well as the eigenvalues ${u_n} 's.$  

\chapter{Discussions}

In this paper, we started with the propositions [A] and [B] so as  to get 
the general framework describing the nonlocalizable object $U$ beyond
the bilocal field
theory and arrived at the noncommutative matrix representation of space-time 
quantities under $U$-field representation.

At this point, it is quite interesting to notice the recent
development in the 
string theory,\ref\sussk which also aims a unified theory of
everything. It is noticeable there that the apparently different
string models investigated so far seem to be equivalently 
connected with each other in terms of {\it duality}, and further
surprisingly space-coordinates appear in the form of noncommutative
and infinite-dimensional matrices, likely in our present theory,
although the string theory 
seems to start with the conventional 
\def\sussk{E.~Witten, Nucl.~Phys.~{\bf B460} (1995), 335.\nextline
T.~Banks, W.~Fischer, S.~H.~Shenker and L.~Susskind, Phys.~Rev.~{\bf D55} 
(1997), 5112.\nextline
M.~Li and T.~Yoneya, Phys.~Rev.~Letters {\bf 78} (1997), 1219.}
space-time concept. 

Therefore, it is quite important to clarify the origin of the
matrix-representation in both approaches. In the string theory, it
appears as $ N$ infinite limit of $N \times N$ matrix, and $N$ denotes
the number of the so-called D(irichlet)0-branes as the fundamental
constituents of the string system. It strongly suggests that the n-th
basis vector in our $U$-field representation, $\vert n \rangle$ (the
n-th excitation mode of nonlocalizable object $U$ characterized by its
eigenvalue $u_n$), in general, corresponds to the various 
composite
state of the infinite number of D0-branes(or D-particles) as the
fundamental constituents of the string system. It is interesting to
conjecture that there exists some kind of transformation which
connects both representations beyond their apparent difference.

Furthermore, in the present paper, the argument of several important
problems remains to be done, such as the second quantization of
$U$-field or the fundamental equation of $U$-field involving
interactions. 

\ack{
The present author would like to thank S. Yahikozawa, K. Itoh and T. Kugo
for the kind information and discussion on the recent development in
string 
theory. The author is deeply indebted  to Prof. S. Ishida for valuable
discussion and constant encouragement of the study of space-time
description of elementary particles.}

\refout
\end